\documentclass[11pt]{book}
\usepackage{a4wide}
\usepackage{fancyheadings}
\usepackage{setspace}
\usepackage{newcent}
\usepackage{xspace}
\usepackage{amssymb}
\usepackage{amstext}

\newcommand{\Edd}{\emph{Eddington}\xspace}
\newcommand{\Cor}{COROT\xspace}

\begin{document}
\thispagestyle{empty}

\begin{center}
  \section*{Planetary transits and stellar variability}
  \subsection*{Suzanne Aigrain}
\end{center}

\vspace{1cm}

\noindent Most of the 130 or so exo-planets (planets orbiting a star
other than the Sun) known to date, were detected via the radial
velocity (RV) method, which relies on the spectroscopic detection of
changes in the parent star's radial velocity as it orbits the
star-planet system's centre of mass. Another promising method relies
on the detection, in stellar photometric time series, of the periodic
dips caused by planets as they cross the disk of their parent star:
planetary transits. 

Many ground-based transit search projects have been operating for
several years and are expected to come to fruition soon -- a handful of
planets detected via their transits have been confirmed already. From
the ground, both transit and RV methods are limited to giant
planets. Several space-based transit search missions are thus planned
to probe the terrestrial and habitable planet regimes. The preparation
of data analysis tools for these missions, in particular \Cor and
\Edd, has been the focus of my PhD, with potential application to
ground-based data as a secondary objective.

I first developed and tested an algorithm for the automated detection
of transits in white noise, a challenge due to the rare, brief and
shallow nature of the transits. One of the most important noise
sources for future space-based missions is the intrinsic low-amplitude
variability of the parent star on timescales of tens of minutes to
weeks. I constructed an empirical model of this `stellar
micro-variability' to simulate realistic light curves for a variety of
stars, and developed filters to remove micro-variability. Monte Carlo
simulations were used to test the performance of these tools alone and
in combination, and to identify which types of stars make the most
promising targets for \Edd \& \Cor. 
 
The algorithms' performance was tested against that of others by
participating in the \Cor transit detection blind exercise, in which a
number of groups from across Europe applied their algorithms to a set
of simulated light curves of content known only to a game master. A
transit search was also performed in 5 nights of data obtained in 2003
by the UNSW transit search team using the 0.5 APT telescope in Siding
Springs Observatory in the field of open cluster NGC 6633. and a
handful of transit candidates with depths below 50 mmag were
identified.

\end{document}